# A Hidden Markov Model Based Unsupervised Algorithm for Sleep/Wake Identification Using Actigraphy

Xinyue Li, PhD, [1,2] Yunting Zhang, PhD, [1,3] Fan Jiang, MD, PhD, [3,4] Hongyu Zhao, PhD [2,5]

**Affiliations**

1. Child Health Advocacy Institute, Shanghai Children's Medical Center, Shanghai Jiao Tong University School of Medicine, Shanghai, China
2. Department of Biostatistics, School of Public Health, Yale University, New Haven, CT, USA
3. School of Public Health, Shanghai Jiao Tong University, Shanghai, China
4. Department of Developmental and Behavioral Pediatrics, Shanghai Children's Medical Center, Shanghai Jiao Tong University School of Medicine, Shanghai, China
5. Shanghai Jiao Tong University – Yale Joint Center for Biostatistics, Shanghai, China

**Address correspondence to:** Jiang Fan, MD, PhD and Hongyu Zhao, PhD

Jiang Fan, MD, PhD
Department of Developmental and Behavioral Pediatrics, Shanghai Children's Medical Center, Shanghai Jiao Tong University School of Medicine, Shanghai, China
Postal address: 1678 Dongfang Road, Pudong, Shanghai, China
Telephone number: 86(21) 38626161
Email: fanjiang@shsmu.edu.cn

Hongyu Zhao, PhD
Department of Biostatistics, Yale School of Public Health, New Haven, CT, USA
Postal address: 60 College Street, Suite 201, New Haven, CT, USA
Telephone number: 1(203)7853613
Email: hongyu.zhao@yale.edu

**Running head:** Automated Sleep/Wake Detection Using Actigraphy

**Total number of words:** 4,843

**Number of references:** 25

**Conflict of Interests:** The authors have no potential conflicts of interest relevant to this article to disclose.

**Abstract**

Actigraphy is widely used in sleep studies but lacks a universal unsupervised algorithm for sleep/wake identification. In this study, we proposed a Hidden Markov Model (HMM) based unsupervised algorithm that can automatically and effectively infer sleep/wake states. It is an individualized data-driven approach that analyzes actigraphy from each individual respectively to learn activity characteristics and further separate sleep and wake states. We used Actiwatch and polysomnography (PSG) data from 43 individuals in the Multi-Ethnic Study of Atherosclerosis to evaluate the performance of our method. Epoch-by-epoch comparisons were made between our HMM algorithm and that embedded in the Actiwatch software (AS). The percent agreement between HMM and PSG was 85.7%, and that between AS and PSG was 84.7%. Positive predictive values for sleep epochs were 85.6% and 84.6% for HMM and AS, respectively, and 95.5% and 85.6% for wake epochs. Both methods have similar performance and tend to overestimate sleep and underestimate wake compared to PSG. Our HMM approach is able to quantify the variability in activity counts that allow us to differentiate relatively active and sedentary individuals: individuals with higher estimated variabilities tend to show more frequent sedentary behaviors. In conclusion, our unsupervised data-driven HMM algorithm achieves slightly better performance compared to the commonly used algorithm in the Actiwatch software. HMM can help expand the application of actigraphy in large-scale studies and in cases where intrusive PSG is hard to acquire or unavailable. In addition, the estimated HMM parameters can characterize individual activity patterns that can be utilized for further analysis.

# Introduction

In sleep studies, it is important to have accurate measurement of sleep duration. Questionnaires and sleep diaries, either self-reported or recorded by others, are commonly used as they are easy to administer and can directly provide the information on sleep start, sleep end and sleep duration. However, such methods may be subjective and have potential bias.(Murphy, 2009, Shephard, 2003) On the other hand, polysomnography (PSG) is considered as the "gold standard" in sleep studies, but has limited use due to its high cost, in-lab setting, intrusive measures, and difficulty in long-time monitoring. For example, it is not easy to use PSG in pediatric populations due to difficulties in wearing invasive sensors and wires and not feasible to use PSG in large-scale epidemiological studies either.(Acebo et al., 2005)

Recently, actigraphy has been adopted in sleep studies as an alternative to sleep diaries and PSG. It uses an accelerometer that works by monitoring acceleration in one or more directions, and this wristwatch-like device is often worn on the wrist to record activity continuously for several days. Either the raw data or the transformed activity count data can be used to study sleep-wake patterns and screen sleep disorders. Actigraphy not only avoids the subjectivity and bias issues with sleep diaries but also overcomes the drawbacks of PSG. While actigraphy does not contain as rich information as PSG, it is useful when long-time and non-invasive monitoring is required.

Several studies have been conducted to evaluate the usefulness of actigraphy in identifying sleep and wake states compared to PSG and sleep diaries.(Cole et al., 1992, Sadeh et al., 1994, Werner et al., 2008, Weiss et al., 2010, Meltzer et al., 2012b, Belanger et al., 2014) The sleep/wake identification algorithm varies depending on the device used, the study population, and the wearing method. Different devices such as Actiwatch and ActiGraph (GTX3) can have different outputs, such as 1-dimension or 3-dimension, transformed activity count data or 50-Hz raw data. Different populations such as infants, adolescents and adults, and elderly people can have very different sleep and activity patterns.(Meltzer et al., 2012a, Meltzer et al., 2012b, Sadeh, 2011) Wearing the device on wrist, ankle or hip also affects the activity data recorded.(Zinkhan et al., 2014) As a result, the sleep/wake identification algorithms are often developed separately for each particular use, and the validation data from PSG or sleep diaries are always required to train the algorithm.

The most commonly used sleep/wake scoring algorithms were developed by Cole et al. (1992) and Sadeh et al. (1994). These algorithms utilize information such as activity counts in the previous, the present, and the following epochs as well as the mean and standard deviation of activity counts in the scored epoch window to build the predictive logistic regression models. Because logistic regression models are easy to build and implement, most studies have adopted this approach and include different variables to develop sleep/wake identification algorithms.(Meltzer et al., 2012a, Sadeh, 2011) A new approach based on artificial neural networks and decision trees was proposed to improve the logistic regression models by taking into account non-linear effects.(Tilmanne et al., 2009) Nonetheless, all these models have the drawbacks that they rely heavily on the dataset that the model is trained on, so the developed algorithm is ad-hoc and might only work for one particular dataset. This largely limits the application of the algorithm, unless the device, the population, and the wearing method of the new dataset are the same as before. Otherwise, one has to collect PSG or sleep diaries again and train models a second time, which is labor-intensive.

Another approach, such as the one embedded in the Actiwatch software, is an unsupervised algorithm that does not necessarily require the model training step. It uses information such as 10 consecutive epochs below a pre-specified immobility threshold as the sleep start and consecutive epochs above a pre-specified mobility threshold as the wake start.(CamNtech Ltd., 2008) The sleep and wake criteria are somehow arbitrary but easy to apply. The major problems with this approach are the choices of thresholds and lengths of windows. The Actiwatch software algorithm is validated against certain populations to achieve desired accuracy, but generalization of the algorithm to other populations needs further validation.(Kushida et al., 2001, CamNtech Ltd., 2008)

In this paper, we propose a sleep/wake identification method based on Hidden Markov Model (HMM) that has several advantages over existing methods. First, it is an unsupervised algorithm that does not require training data such as PSG and sleep logs to train the model. Second, it can be directly applied to datasets from different devices and populations, as it is data-driven that makes full use of the information contained in the dataset to learn and separate sleep and wake states. Third, unlike the Actiwatch software algorithm, it does not use subjective thresholds, which are difficult to choose and justify. We note that HMM has been widely used in pattern recognition and biological sequence analysis, and it is easy to implement and computationally efficient.(Nasrabadi, 2007, Yoon, 2009) Actigraphy coupled with the HMM algorithm makes it feasible to apply actigraphy in large epidemiological studies and also serves as an alternative to PSG for multiple-night monitoring in a natural environment.(Acebo et al., 2005, Malow et al., 2006, Meltzer et al., 2012a, Meltzer et al., 2012b)

The rest of this article is organized as follows. In the Methods section, we discuss the proposed HMM algorithm in detail, and we also describe the algorithm implemented in the Actiwatch software. We then introduce the Multi-Ethnic Study of Atherosclerosis (MESA) sleep data that we use to implement and evaluate the HMM algorithm. In the Results section, we present the HMM results and comparison with the Actiwatch software algorithm. In the Discussion section, we conclude with the implications of our results and future research directions.

## Methods

### Hidden Markov Model (HMM)

Our proposed unsupervised algorithm for sleep/wake identification is based on a two-state Hidden Markov Model (HMM).(Baum and Petrie, 1966, Baum et al., 1970) In the model, we assume that the sequence of the observed activity counts are generated from an unobservable two-state Markov chain, with the two states being sleep and wake. In the sleep state, the activity counts are mostly zeros with some low activity counts, while in the wake state, the activity counts are generally high with some low counts denoting sedentary behaviors. Therefore, we assume that the activity counts follow different distributions under the sleep and wake states, and we can infer the hidden sleep/wake states at each time point based on the observed count data.

In our model, we consider the log transformed data: log(count+1) as the observed data. Although we can directly model the activity count using a Poisson or Negative Binomial distribution in the wake state and a zero-inflated Poisson distribution in the sleep state, the observed activity counts can range from 0 to 4,000 per epoch, and this large range poses both statistical and computational challenges in data analysis. Therefore, we choose to model the log transformed count data and our empirical results suggest that the HMM algorithm works well for the log transformed data.

The structure of our proposed HMM model is shown in Figure 1. We observe activity count data from time 1 to time $T$: $O^{(T)} = \{O_1, O_2, ..., O_T\}$, where $O_i$ denotes the log transformed activity count in the $i$th epoch. Let $X^{(T)} = \{X_1, X_2, ..., X_T\}$ denote the sequence of the corresponding hidden states across these T time points, where each $X_i$ can be one of the two possible hidden states $S = \{s_1, s_2\}$ in each epoch, with $s_1$ denoting the sleep state and $s_2$ denoting the wake state. Thus, $X_t \in S = \{s_1, s_2\}$, $t = 1, 2, ..., T$.

(Figure 1 goes here)

We assume that $X_i$ follows a Markov model, that is the hidden state $X_{t+1}$ at time $t+1$ solely depends on $X_t$, and the observation $O_t$ at time $t$ solely depends on the hidden state $X_t$:

$$P(X_{t+1}|X^{(t)}) = P(X_{t+1}|X_t)$$

$$P(O_{t+1}|X^{(t+1)}, O^{(t+1)}) = P(O_{t+1}|X_{t+1})$$

$A$ in Figure 1 denotes the transition probability. In our case, the transition matrix is a 2 by 2 matrix, in which $a_{ij}$ represents the transition probability from state $s_i$ to state $s_j$:

$$P(X_{t+1} = s_j | X_t = s_i) = a_{ij}, \quad s_i, s_j \in S$$

The emission probability $P(O_t|X_t)$ denoted by $B$ in Figure 1 depends on the state of $X_t$.

If $X_t = s_1$ in the sleep state, we assume that the log transformed count follows zero-inflated truncated Gaussian distribution, which is truncated from 0 to the left. It has a zero component because sleep is associated with rare movements and activity measurements during sleep often involve many zeros. Therefore:

$$b_1(0) = P(O_t = 0 | X_t = s_1) = \alpha + (1-\alpha) \cdot \frac{\frac{1}{\sigma_1}\phi\left(\frac{0-\mu_1}{\sigma_1}\right)}{1 - \Phi\left(\frac{0-\mu_1}{\sigma_1}\right)}$$

$$b_1(k) = P(O_t = k | X_t = s_1) = (1-\alpha) \cdot \frac{\frac{1}{\sigma_1}\phi\left(\frac{k-\mu_1}{\sigma_1}\right)}{1 - \Phi\left(\frac{0-\mu_1}{\sigma_1}\right)}$$

where $\alpha$ is the probability of extra zeros, $\mu_1$ is the mean, $\sigma_1$ is the standard deviation, $\phi(\cdot)$ is the probability density function of the standard normal distribution, and $\Phi(\cdot)$ is its cumulative distribution function.

If $X_t = s_2$ in the wake state, we assume that the log transformed count follows the Gaussian distribution:

$$b_2(k) = P(O_t = k \mid X_t = s_2) = \frac{1}{\sigma_2}\phi(\frac{k - \mu_2}{\sigma_2})$$

where $\mu_2$ is the mean and $\sigma_2$ is the standard deviation of the Gaussian distribution.

Therefore, the set of parameters for the emission probability is $B = \{B_1, B_2\} = \{\alpha, \mu_1, \sigma_1, \mu_2, \sigma_2\}$. To initiate the Markov chain, we also need the initial state probabilities $\Pi = \{\pi_0, \pi_1\}$ that denote the probability of being in the sleep or the wake state at time $t = 1$, respectively. Given the transition probabilities, emission probabilities, and initial state probabilities $\Theta = \{A, B, \Pi\}$, HMM can be fully specified.

To obtain $\Theta^* = \text{argmax}_\Theta P\{O^{(T)} \mid \Theta\}$, we can use the Baum-Welch algorithm, which employs the expectation-maximization algorithm to find $\Theta^*$ that maximizes $P\{O^{(T)} \mid \Theta\}$, namely the probability of observing the sequence of count data.(Baum et al., 1970) Then based on the estimated $\Theta^*$, we further look for the optimal path of hidden states $X^{(T)^*} = \text{argmax}_{X^{(T)}} P\{X^{(T)}, O^{(T)} \mid \Theta^*\}$ using the Viterbi algorithm, a dynamic programming method.(Ryan and Nudd, 1973) The optimal hidden states $X^{(T)^*}$ are exactly the sequence of inferred sleep/wake states. Based on the obtained sequence of hidden states, we focus on the same-state sequences longer than 15 minutes and smooth out shorter sequences to ensure that it captures stable sleep durations.

Our proposed HMM is similar to the Actiwatch software algorithm in that it uses the activity counts (low/high) to infer which state, sleep or wake, most likely generated the observed activity counts. The HMM is able to suppress frequent transitions between two states, yielding relatively smooth results, namely sequences of stable consecutive wake and sleep states. The model assumptions on the zero-inflated truncated Gaussian and Gaussian for the respective sleep and wake states also work well, or otherwise the algorithm will not converge. The normality assumptions for the two states are assessed through the density plots and QQ-plots. Processing of Actiwatch data and implementation of the HMM-based algorithm is in R (version 3.6.2).

**Other Methods**

In the following, we will compare our HMM algorithm with other methods, including the Actiwatch software algorithm. Note that our method is an unsupervised algorithm, which does not require labeled outcomes, such as true sleep/wake states in our scenario, to train the model and obtain results. In comparison, popular methods such as the logistic regression algorithm in Cole et al. (1992) are supervised algorithms that require labeled outcome upfront, without which it is impossible to train the model.(Cole et al., 1992) Therefore, we only compare our results to other unsupervised algorithms such as the one in the Actiwatch software to evaluate the performance of HMM.

Our description of the Actiwatch Software (AS) algorithm is based on the Actiwatch manual and also the MESA sleep data documentation guide that provides the sleep/wake detection algorithm.(CamNtech Ltd., 2008, Zhang et al., 2018, Dean et al., 2016) First, it requires the input of go-to-bed time and get-up time, namely the two time points recorded in the sleep diary. Second, it re-scores each data point from each epoch and those surrounding it to make a total score. Specifically, the adjacent epoch within 1 minute is reduced by a factor of 5 and added to the current epoch, and the adjacent epoch within 2 minutes is reduced by a factor of 25 and also added to obtain the total score. Then, to determine the sleep start, the algorithm starts from the go-to-bed time to look for a period of 10-minute consecutive epochs below the immobility threshold (4 counts per minute) allowing for 1-minute epoch above the threshold, and the start of this period is considered as the sleep start. To determine the sleep end, it looks backwards from the get-up time for consecutive 6-minute epochs below the threshold (6 counts per minute) allowing 2 epochs above the threshold, and the last epoch of the period is considered as the sleep end. In short, the Actiwatch software algorithm is an unsupervised algorithm that does not require training data or model fitting but only threshold setting in order for the algorithm to work. However, the selections of thresholds and the choice of the length of windows for determining sleep/wake states are somewhat subjective and might need justification and validation.

**Method Comparisons**

To compare the HMM algorithm with the AS algorithm, epoch-by-epoch analysis with polysomnography (PSG) as the reference standard was performed. Accuracy or percent agreement is calculated as the total number of epochs that were correctly classified by the algorithm divided by the total number of epochs. Sensitivity for sleep is calculated as the proportion of epochs PSG-scored sleep epochs that are correctly classified as sleep by HMM or AS. Specificity for sleep is calculated as the proportion of epochs PSG-scored wake epochs that are correctly classified as wake by HMM or AS. Positive predictive value for sleep is calculated as the proportion of actigraphy-scored sleep epochs that are scored the same by PSG, and positive predictive value for wake is calculated in the same manner.

Sleep variables are also calculated based on PSG scored sleep epochs as well as HMM and AS scored sleep epochs from actigraphy and then compared. Total epochs scored are the same across the three methods, as we only used matched PSG and actigraph data. We computed total sleep time, sleep latency, wake time after sleep onset (WASO), and sleep efficiency. Sleep latency is defined as the time from lights out to first epoch of sleep. Sleep efficiency is defined as total sleep time divided by total time from lights out to lights on. These sleep variables were computed based on each individual data, and mean and standard deviation were reported. Pearson correlations and paired t-tests were used to compare between PSG and HMM and between PSG and AS on each of the sleep variables.

**Data**

Data were from the Multi-Ethnic Study of Atherosclerosis (MESA), and its study design has been published before.(Zhang et al., 2018, Dean et al., 2016) Briefly, MESA is a multisite prospective study for cardiovascular diseases. In 2000-2002, 6,814 participants who aged 45-84, identified themselves as white, African American, Hispanic or Chinese American and were free of clinically apparent CVD enrolled in the study from six US communities. At MESA Exam 5 in 2010-2013, 2,261 individuals participated in the sleep exam (59.7%), which included full overnight PSG, 7-day wrist-worn actigraphy, and a sleep questionnaire in 2010-2012.(Zhang et al., 2018, Dean et al., 2016) Institutional Review Board approval was obtained at each site and written informed consent was obtained from all participants. We used one night PSG data and the corresponding actigraph data for algorithm validation. While there are over two thousand participants in the study, we only selected data with high quality scores, using stringent criteria of actigraphy quality scored as 7 and PSG quality scored as 6 or 7. An actigraphy quality score of 7 suggests reliable overnight data and event markers including bed time consistent with sleep diaries.(Zhang et al., 2018, Dean et al., 2016) The PSG quality score measures the total duration of useable and artefact-free signals across channels that are critical for accurate sleep scoring, and the documentation suggests using data with PSG quality scored 5 or above.(Zhang et al., 2018, Dean et al., 2016) After quality filtering, data from 43 individuals were used in algorithm validation, including 41,348 30-second epochs of actigraph data and corresponding PSG data.

**Results**

The characteristics of the 43 individuals are shown in Table 1. Four individuals had sleep apnea and one had restless leg syndrome as diagnosed by the doctor. Other individuals did not report any sleep-related conditions. The Epworth Sleepiness Scale measures the propensity to fall asleep in certain situations.(Johns, 1991) It has a range between 0-24 with a score $\geq$ 10 indicative of excessive daytime sleepiness. Five individuals had scores higher than fifteen that suggested higher sleep propensities.

(Table 1 goes here)

**HMM Parameters**

As shown in Table 2, the estimated percentage of zeros in the sleep state has mean 0.73. In the truncated Gaussian part for the sleep state, $\hat{\mu}_1$ has mean 2.49 and a range from 1.75 to 3.07, which corresponds to 12 activity counts per epoch and a range from 6 to 22 activity counts per epoch on the original scale. The estimated standard deviation for the truncated Gaussian has mean 1.25, and the range from 1.13 to 1.50 is relatively narrow. For the wake state, $\hat{\mu}_2$ has mean 4.80 and ranges from 3.97 to 5.41, which corresponds to 122 activity counts per epoch and a range from 53 to 224 activity counts per epoch on the original scale. The estimated standard deviation has mean 0.87, and it has a wider range from 0.64 to 1.17 compared to the sleep state. A smaller standard deviation suggests more concentrated activity counts around the mean while a larger standard deviation suggests a more spread-out pattern.

(Table 2 goes here)

Among the five HMM parameters, the estimated standard deviation for the wake state has a relatively large variation, which provides us with insights into the physical activity variability across individuals. Figure 2 shows the activity plots for ID 3396 and ID 4729. Both individuals are white females with similar age without sleep-related conditions such as sleep apnea, restless leg syndrome or insomnia. Their mean activity levels during sleep are similar, 8.2 and 9.5 activity counts per 30-second epoch as estimated by HMM. However in the wake state, the estimated standard deviations for these two individuals suggest different activity patterns. ID 3396 has a large $\hat{\sigma}_2$=1.14 and a wider span of activity counts in the wake states that suggest relatively more activity variability and more sedentary behaviors. In comparison, ID 4729 has a smaller $\hat{\sigma}_2$=0.65 and the activity counts are centered around the mean with few low counts for sedentary behaviors. The activity patterns of the two individuals differ in activity levels, activity variability and sedentary behavior tendencies, the information of which is captured by HMM estimates.

(Figure 2 goes here)

The estimated transition matrix from HMM is as expected, as shown in Table 2: the probability for the next epoch to stay in the same state as the current epoch is about 95%-96%, namely that the next epoch is much more likely to stay in the same state.

The normality assumptions for the two states can be studied through the density plots and QQ-plots as shown in Figure 3. Since the sleep state contains a zero component and a truncated normal component, we simulated truncated normal distribution to generate the QQ plot. Overall, the normality assumption seems reasonable as suggested by the QQ plot, though the two distributions may have longer and heavier tails to accommodate occasional movements during sleep and sedentary behaviors during wake. The distribution for the wake state for some individuals has a wider span that corresponds to a larger standard deviation. The distribution for the wake state is slightly skewed to the left, possibly due to sedentary behaviors giving low activity counts.

(Figure 3 goes here)

**Comparison with Actiwatch Software Algorithm**

As shown in Table 3, the overall agreement/accuracy of HMM and AS scored sleep/wake epochs with PSG-scored epochs is about the same, 85.7% and 84.7%, with HMM having a slightly better performance. The specificity is high for both methods, over 97% for all individuals. However, the sensitivity is low for both methods, having low probabilities of correctly predicting wake epochs during sleep. The predictive value of sleep, which means that given the sleep epoch scored by HMM or AS, the probability that it is a true sleep epoch as scored by PSG, is high for both methods, 85.6% and 84.6% for HMM and AS respectively. The predictive value of wake is higher for HMM (95.5%) compared to AS (85.6%).

(Table 3 goes here)

Comparison of sleep variables computed from PSG scored epochs and actigraph scored epochs by HMM and AS is shown in Table 4. Paired t-tests comparing HMM and PSG and comparing AS and PSG estimated sleep variables are all significant (p-value < 0.05). Compared to PSG, the total sleep time estimated by both HMM and AS is longer, with AS the longest, and the sleep efficiency estimated by HMM and AS is thus higher as well. Sleep latency and WASO are shorter for both HMM and AS. Both algorithms tend to overestimate sleep epochs and underestimate wake epochs. For all sleep variables, the mean difference between the PSG scored variable and the HMM scored variable was smaller than the mean difference between the PSG scored variable and AS. Overall, HMM has better performances than AS.

(Table 4 goes here)

**Discussion**

**HMM Algorithm**

Overall, the HMM algorithm works well under the Gaussian distribution assumption for the wake state and the zero-inflated truncated Gaussian distribution for the sleep state for the log transformed activity count. The Gaussian distribution in the wake state is left-skewed to include sedentary behaviors, and the truncated Gaussian distribution component in the sleep state has slightly heavier tails to accommodate rare movements during sleep. Frequent transitions between sleep and wake states are discouraged by the transition matrix, which gives a higher probability to stay in the current state and serves as the penalty to suppress transitions. The mechanism for HMM is similar to the Actiwatch algorithm, as it considers sequences of low/high activity counts to infer sleep/wake states, though a little implicitly, in the algorithm. In summary, the two activity count distributions and the transition between sleep and wake states are well captured by HMM.

The HMM algorithm is also computationally efficient that enables fast processing of large actigraphy datasets. It has been extensively applied to biological sequence analysis such as RNA analysis, in which the lengths of sequences are much longer.(Yoon, 2009) In our analysis, processing one-day actigraphy or seven-day actigraphy of over 20,000 epochs finishes in seconds. High computational efficiency allows for fast processing of actigraphy in large-scale epidemiological studies.

**Comparison with the Actiwatch Software Algorithm**

To evaluate the performance of HMM, we compare it with the algorithm embedded in Actiwatch software that is also an unsupervised algorithm. HMM has slightly better performance in terms of accuracy and sensitivity as well as predictive value of sleep and predictive value of wake. Both methods tend to overestimate sleep epochs and underestimate wake epochs with PSG as the reference. This is consistent with actigraphy validation studies in the literature. In the actigraphy reliability study by Marino et al. (2013), it reported a mean specificity of 33%, and the specificity is lower for individuals with lower sleep efficiency. In another study by Meltzer et al. (2012b), specificity is higher among healthy younger individuals with higher sleep efficiency. Because our study subjects are in the middle to elderly population, the specificity is not high. The overall accuracy (85.7%) for the HMM algorithm is high and the sensitivity to identify sleep epochs is above 97% for each participant. Thus we conclude that the HMM algorithm based on actigraphy is a useful and valid means to estimate total sleep time, with some limitations in specificity. In our study, the HMM outperforms the Actiwatch software algorithm with closer sleep estimates to PSG data.

The Actiwatch software has the advantage that it takes information such as sleep latency into consideration. However, the subjective activity count thresholds and fixed length of windows may not work for all populations and one always needs validation against a group of individuals to achieve optimal performance. Another limitation is that an ad-hoc algorithm like the Actiwatch software algorithm may not consider all unique/specific situations in data. For example, if the first ten epochs for sedentary behaviors meet the criteria to be considered as sleep mistakenly, then even if the eleventh epoch has high activity counts, it will still be ignored but treated as movement during sleep. It is not easy to consider all such situations and build a "perfect" ad-hoc algorithm.

In comparison, HMM can be directly applied to data from different populations or difference devices as it is data-driven that fully utilizes the information contained in data to learn individual activity characteristics and further distinguish sleep/wake states. We note that HMM parameters can capture heterogeneities in activity patterns: some individuals are more active, showing relatively more concentrated high activity counts during the day with few sedentary behaviors (relatively small σ values estimated for the wake state), while some tend to have more sedentary behaviors and wider spans of activity counts during the day (relatively large σ values). Differences in the estimated HMM parameters can be used as individual sleep or activity features in further analyses.

**Comparison with Supervised Methods**

For supervised methods such as logistic regression, random forest, and decision trees, they require labeling of the sleep/wake states (either PSG or sleep logs) in order to train the model.(Cole et al., 1992, Sadeh et al., 1994, Tilmanne et al., 2009) This requirement largely limits the use of supervised algorithms when only actigraph data are available. For example, in pediatric sleep studies, PSG data are hard to obtain from young children due to difficulties in wearing invasive sensors and wire.(Yang et al., 2014) In the example of large-scale epidemiological studies planning to monitor sleep for a period of time, collecting PSG is not feasible as it is often conducted in lab settings and limited to small sizes, typically under a hundred. Obtaining high-quality sleep logs in large-scale studies is labor-intensive and likely infeasible. In these cases, supervised algorithms often do not work. In comparison, unsupervised methods such as HMM can be applied.

In addition, supervised methods rely heavily on the data used for training the model, which limits the use of the algorithm to a few datasets but not others. If new data are from a different population or a different device, the existing algorithm cannot be applied to the new data with confidence and new model building and training is required. In comparison, regardless of the new data source and whether outcome labels are available or not, the data-driven HMM algorithm can be directly applied to identify sleep/wake states efficiently.

On the other hand, HMM can also be used as a supervised method to train models and infer sleep/wake states. Our HMM algorithm consists of two steps: first, estimate the model parameters, and second, infer the sequence of hidden sleep/wake states. If we have labeled sleep/wake epochs, then instead of using the Baum-Welch algorithm in the first step, we can directly estimate model parameters based on the labeled two states and jump to the second step to infer sleep/wake states. In this way we can leverage prior knowledge on sleep/wake states to train the algorithm.

It is noteworthy that our proposed HMM unsupervised algorithm considers individual variabilities in sleep/wake identification that existing methods do not take into account. Different population can have different sleep and activity characteristics, and within the same population, individual variation also exists. Because HMM analyzes actigraphy from each individual separately, it learns individual characteristics and identifies sleep/wake states accordingly. As an added benefit, HMM parameter estimates that capture individual variabilities in sleep/wake patterns can be utilized in further analysis.

**Limitation and Future Work**

We proposed an unsupervised sleep/wake identification algorithm based on actigraphy and evaluated the performance using data from the MESA study. The study subjects included individuals aged 56-87 of different races/ethnicities, which represents a diverse population in the middle to elderly age, and future work may further validate the algorithm among the younger population. It is noteworthy that the HMM algorithm is data-driven that fully utilizes the information in data rather than using a uniform criterion, and thus applying it to other populations will automatically learn population activity characteristics effectively and identify sleep/wake epochs accordingly. Another limitation of our study is that the specificity to detect wake epochs during sleep is not high. This problem also exists in other actigraphy studies in literature, as the gold standard PSG may contain more information that can recognize wake epochs during sleep. As wearable devices have been undergoing rapid development to include more measurements, future work can consider incorporating metrics such as heart rates, body temperature and electrocardiogram, which exhibit different patterns during sleep, to refine HMM. It is easy to extend HMM to multi-dimensional data and improve algorithm performance.

**Summary**

Our proposed HMM algorithm is effective in inferring sleep/wake states based on actigraphy and outperforms the Actiwatch software algorithm in our study. It has many advantages. First, compared to supervised methods, our HMM approach is unsupervised and can save much manual work in training data collection, model selection, and model fitting. It can largely expand the application of actigraphy when outcome records, PSG or sleep logs, are not available. Second, it can be directly applied to data from different populations or wearable devices, since the HMM algorithm is data-driven that makes full use of the information contained in the dataset to separate sleep and wake states. Third, it considers individual variations in sleep identification that none of the existing methods consider. The estimated HMM parameters are individual-specific and informative on individual activity patterns, the features of which can be leveraged in further analyses. Given the easy implementation and good performance of the HMM algorithm, it can be widely applied in clinical research and aid in the use of actigraphy in large-scale epidemiological studies.

**Figures and Tables**

Figure 1. Left panel: an example of an one-day activity plot, with s1 and s2 denoting the sleep state and the wake state respectively. Right panel: a directed graph showing the HMM structure.

Figure 2. Activity plots for ID 3396 and ID 4729, in which HMM estimated parameters suggest individual variabilities in activity patterns. ID 3396 is a white female aged 65 and ID 4729 is a white female aged 60, both without sleep-related conditions. Red line: smoothed curve of log activity counts using locally-weighted polynomial regression.

Figure 3. Density plots and Q-Q plots for the truncated normality assumption in the sleep state and normality assumption in the wake state respectively.

Table 1. Characteristics of subjects from the MESA study.

Table 2. Estimated HMM parameters based on actigraph data from 43 individuals.

Table 3. Comparison between the HMM algorithm performance and the AS algorithm performance in sleep-wake detection by validating against polysomnography.

Table 4. Comparison of sleep variables scored by PSG, the HMM algorithm and the AS algorithm based on actigraphy.

Table 1. Characteristics of subjects from the MESA study.

| Variables | Subjects (N=43) |
|---|---|
| Gender, n (%) | |
|   Male | 21 (48.8) |
|   Female | 22 (51.2) |
| Race, n (%) | |
|   White | 28 (65.1) |
|   Chinese American | 3 (7.0) |
|   African American | 8 (18.6) |
|   Hispanic | 4 (9.3) |
| Diagnosis by Doctor, n (%) | |
|   Sleep Apnea | 4 (9.3) |
|   Restless legs syndrome | 1 (2.3) |
|   Insomnia | 0 (0) |
| Age, mean (range) | 69 (range=56-87) |
| Epworth Sleepiness Scale Score, mean (SD) | 6.3 (SD=4.7) |

Table 2. Estimated HMM parameters based on actigraph data from 43 individuals.

|  | Parameter | Mean | [Min, Max] |
|---|---|---|---|
| **Sleep** | $\alpha_1$ | 0.731 | [0.529,0.890] |
|  | $\mu_1$ | 2.486 | [1.753,3.072] |
|  | $\sigma_1$ | 1.248 | [1.125,1.497] |
| **Wake** | $\mu_2$ | 4.803 | [3.971,5.406] |
|  | $\sigma_2$ | 0.866 | [0.640,1.165] |
| **Transition** | $a_{11}$ | 0.960 | [0.920,0.979] |
|  | $a_{12}$ | 0.055 | [0.024,0.097] |
|  | $a_{21}$ | 0.040 | [0.021,0.080] |
|  | $a_{22}$ | 0.945 | [0.903,0.976] |

*In the sleep state, log (activity counts + 1) follows zero-inflated truncated Gaussian distribution, with $(\alpha_1, \mu_1, \sigma_1)$ denoting the zero component, mean, and standard deviation respectively; in the wake state, it follows Gaussian distribution, with $(\mu_2, \sigma_2)$ denoting the mean and standard deviation respectively.

Table 3. Comparison between the HMM algorithm performance and the AS algorithm performance in sleep-wake detection by validating against polysomnography.

| Algorithm | Agreement Rate | Correct Prediction of Sleep (Specificity) | Correct Prediction of Wake (Sensitivity) | Predictive Value of Sleep | Predictive Value of Wake |
|---|---|---|---|---|---|
| HMM | 0.857 (0.736-0.961) | 0.993 (0.971-1.000) | 0.364 (0.274-0.542) | 0.856 (0.726-0.960) | 0.955 (0.760-1.000) |
| AS | 0.847 (0.677-0.970) | 0.997 (0.970-1.000) | 0.300 (0.222-0.546) | 0.846 (0.677-0.969) | 0.856 (0.578-1.000) |

[*] Statistics were first computed for each participant; mean (range) is shown in table.

Table 4. Comparison of sleep variables scored by PSG, the HMM algorithm and the AS algorithm based on actigraphy.

| Variable (N=43) | PSG | HMM | | AS | |
| --- | --- | --- | --- | --- | --- |
| | Mean (SD [a]) | Mean (SD) | r [b] | Mean (SD) | r |
| Total Epochs Scored (minutes) | 480.8 (66.8) | 480.8 (66.8) | - | 480.8 (66.8) | - |
| Total Sleep Time (minutes) | 368.4 (56.6) | 408.1 (64.6) | 0.872** | 413.6 (67.2) | 0.864** |
| Sleep Latency [b] (minutes) | 29.2 (29.8) | 23.4 (28.9) | 0.956** | 14.9 (31.0) | 0.890** |
| WASO [c] (minutes) | 38.3 (20.0) | 23.0 (19.2) | 0.447* | 22.8 (23.1) | 0.309 |
| Sleep Efficiency [d] (%) | 90.7 (6.4) | 94.0 (6.7) | 0.822** | 96.0 (7.8) | 0.670** |

[a] SD: standard deviation
[b] Pearson's correlation between measurements from PSG and actigraph scored by HMM or AS. Asterisk indicates significant correlations.
[c] Sleep latency: time from lights out to first epoch of sleep.
[d] WASO: wake time after sleep onset.
[e] Sleep efficiency: total sleep time divided by total time from lights out to lights on.
* p<0.05 ** p<0.001

Figure 1. Left panel: an example of an one-day activity plot, with s1 and s2 denoting the sleep state and the wake state respectively. Right panel: a directed graph showing the HMM structure.

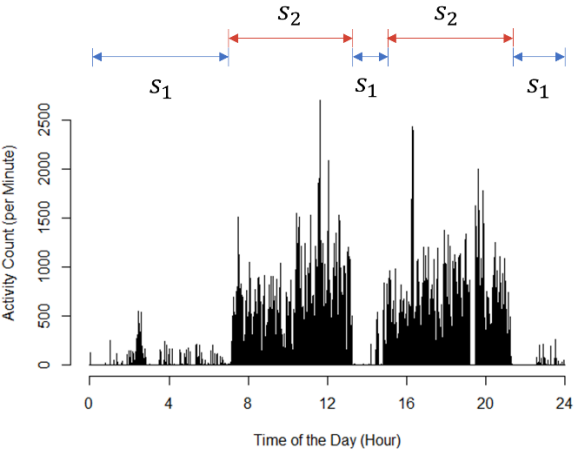
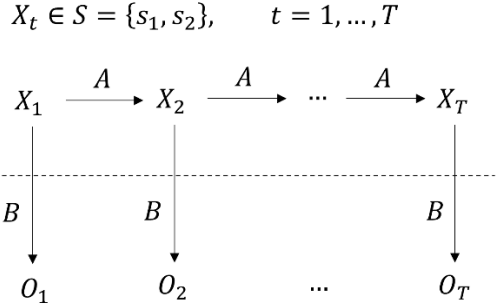

Figure 2. Activity plots for ID 3396 and ID 4729, in which HMM estimated parameters suggest individual variabilities in activity patterns. ID 3396 is a white female aged 65 and ID 4729 is a white female aged 60, both without sleep-related conditions. Red line: smoothed curve of log activity counts using locally-weighted polynomial regression.

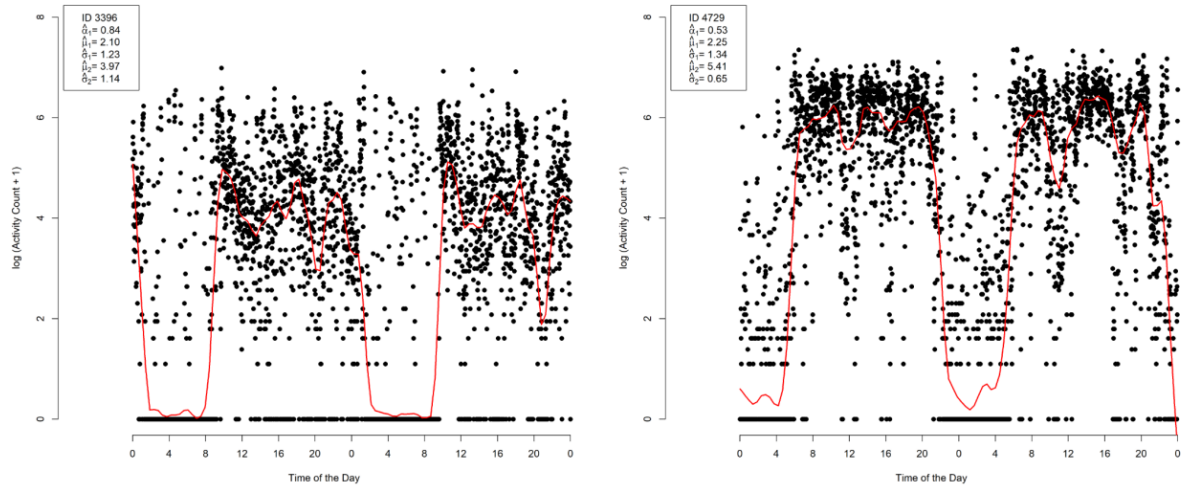

Figure 3. Density plots and Q-Q plots for the truncated normality assumption in the sleep state and normality assumption in the wake state respectively.

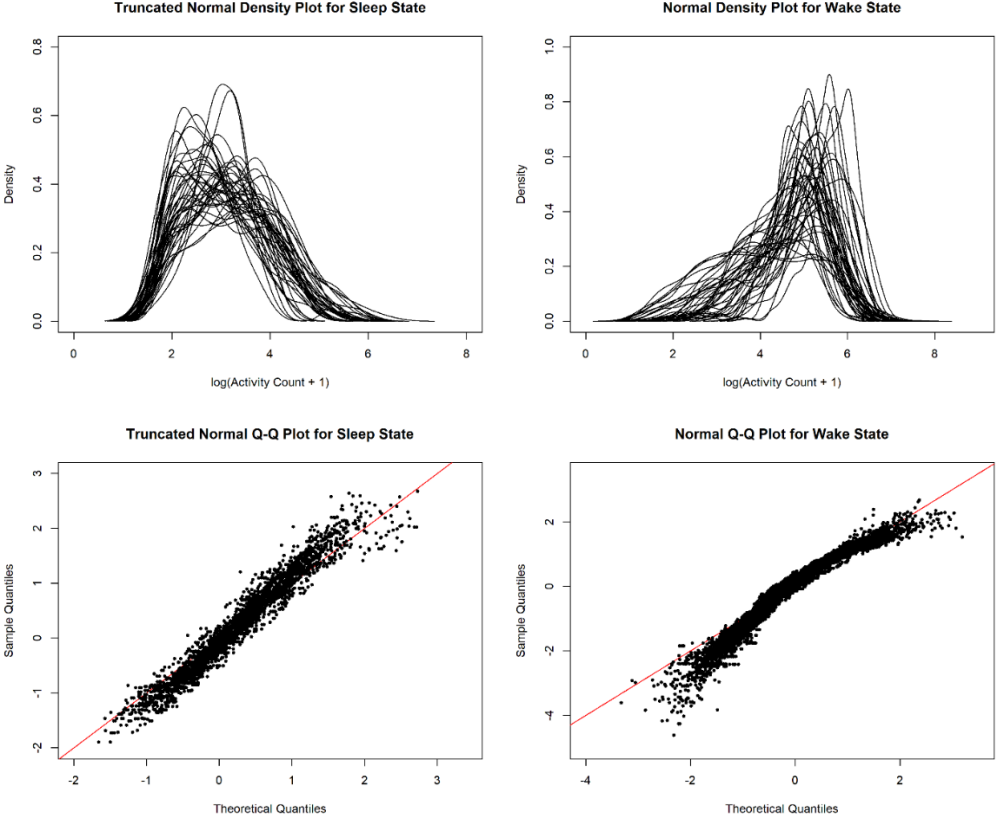